\documentclass[prb,reprint,aps,superscriptaddress,longbibliography]{revtex4-2}

\usepackage{graphicx}
\usepackage{dcolumn}
\usepackage{bm}
\usepackage{braket}
\usepackage{amsmath}
\usepackage{amssymb}
\usepackage{feynmp}
\usepackage{graphics}
\usepackage{enumerate}
\usepackage{color}
\usepackage[hidelinks]{hyperref}
\usepackage[normalem]{ulem} 

\begin{document}


\title{Magnetism and Quantum Melting in Moiré-Material Wigner Crystals}

\author{Nicol\'as Morales-Dur\'an} 
\affiliation{Department of Physics, The University of Texas at Austin, Austin, Texas, 78712, USA}
\author{Pawel Potasz}
\affiliation{Institute of Physics, Faculty of Physics, Astronomy and Informatics, Nicolaus Copernicus University, Grudziadzka 5, 87-100 Toru\'n, Poland}
\author{Allan H. MacDonald}
\affiliation{Department of Physics, The University of Texas at Austin, Austin, Texas, 78712, USA}




\date{\today}
\begin{abstract}
Recent experiments have established that semiconductor-based moiré materials 
can host incompressible states at a series of fractional moiré-miniband fillings. These states have been identified as generalized Wigner crystals in which electrons localize on a subset of the available triangular-lattice 
moiré superlattice sites.  In this article, we use momentum-space exact diagonalization to investigate the many-body ground state evolution at rational fillings from the weak-hopping classical lattice gas limit, 
in which only spin degrees-of-freedom are active at low energies,
to the strong-hopping metallic regime where the Wigner crystals melt.  We specifically address the nature of the magnetic ground states of
the generalized Wigner crystals at fillings $\nu=1/3$ and $\nu=2/3$.

\end{abstract}

\pacs{Valid PACS appear here}
\maketitle

\section{Introduction}
It is now several years since Wu {\it et al.} \cite{FengchengHubbard} 
pointed out that the Hamiltonian of interacting holes in the moiré bands of transition metal dichalcogenide (TMD) heterobilayers can be mapped to 
the triangular lattice Hubbard model. 
Experiments quickly confirmed the validity of this assertion 
by observing Mott insulating states at band filling $\nu \equiv N/M=1$ of the moiré superlattice \cite{HubbardCornell,HubbardColumbia,HubbardETH,Berkeley},
where $N$ is the number of holes and $M$ 
the number of moir\'e unit cells in the system. 
Subsequent experiments have established that TMD-based moiré materials also
exhibit correlated insulating states at a discrete series of fractional fillings of the lowest moir\'e miniband  \cite{CornellWigner,CaliforniaWigner,Berkeley,CornellWignerStripe,STM_Wigner,ContinuousWigner}. 
These insulating states form because electrons localize on a subset of moiré sites in order to minimize strong long-range Coulomb interactions.  Because they break translational symmetry, they are 
reminiscent of the Wigner crystals expected to appear in the two-dimensional electron gas (2DEG) at very low densities \cite{OriginalWigner}. There are however some qualitative differences between Wigner crystals formed in an electron gas
with continuous translational symmetry, and the incompressible states at fractional fillings in moir\'e materials, which have only discrete translational symmetry.  Most importantly, the moiré superlattice potential narrows bands and reduces the relevant single-particle energy scales, making interactions dominant in much of the available phase space.  

The incompressible states in moir\'e superlattices 
are commonly referred to as generalized Wigner crystals and we adopt that
terminology in this paper. The ubiquity of robust crystalline states at fractional fillings in the moiré material platform opens up a new thread in the study of strongly interacting electrons in low dimensions and promises to reveal new physics.  Given
the abundance of distinct moiré semiconductor heterostructures,
even within the group VI transition metal 
dichalcogenide family alone, 
and the ability to tune samples through large ranges of filling factor by varying gate voltage,
it seems likely that it will
be possible to realize a rich variety of generalized Wigner crystal states 
with distinct structural and magnetic properties in the coming years. 

The emergence of incompressible states at non-integer partial band filling 
can be explained only if inter-site electron-electron interactions are included. 
Recent experiments have therefore established moiré TMDs as a platform to simulate extended Hubbard models whose Hamiltonians have tunable hoppings $t_n$, on-site interaction $U_0$, and long-range interaction strengths $U_n$ ($n$ stands for $n$-th neighbor). Assisted hopping and direct exchange non-local interaction terms can also play a crucial role \cite{Nonlocal} in determining the magnetic properties of moiré Hubbard systems. 
The mapping to a Hubbard model is a one-band approximation, whose applicability at  $\nu\le 1$ has generally been confirmed by experiment. 
For fillings above half-filling, there is a competition between the upper Hubbard band and 
remote bands; hence the simple one-band Hubbard model is often insufficient. 
For that reason, in this work we focus on the regime $\nu<1$, having addressed $\nu=1$ in a previous 
study \cite{MoireMIT,Nonlocal}. 

In moiré superlattices, localization of electrons in an insulating state is expected \cite{FengchengHubbard} in the long-moiré-period narrow-moir\'e-band limit. In this regime, the dominant energy scale is $U_0$ at $\nu=1$ and 
$U_1$ for $1/3 \le \nu < 1$. When the twist angle is increased and 
the moiré period decreased, or a displacement field is applied to decrease 
the moiré potential strength, the effective 
hopping parameters $t$ between moiré lattice sites increase and 
eventually become comparable to inter-site interaction strengths $U_1$, complicating the electronic properties.  The interplay between spin and charge degrees of freedom can give rise to different magnetic orders at each filling factor. For example, recent experiments have reported that some of the crystal states are striped phases \cite{CornellWignerStripe}, and that antiferromagnetic interactions are frustrated for $\nu=2/3$ band filling factor \cite{FrustratedWignerMott}. When hopping is strong enough to overcome the near-neighbor interaction, the Wigner crystal will melt into a liquid state -- the Mott-Wigner transition. Interestingly a recent experiment performed on MoSe$_2$/WS$_2$ observed that the charge gap continuously vanishes as the superlattice potential is weakened \cite{ContinuousWigner}. Further experiments have shown that the charge gaps of the generalized Wigner crystal states are asymmetric with respect to half-filling ($\nu=1$) of a single spinful band,
and also with respect to quarter-filling ($\nu=1/2$) \cite{CornellWigner} and demonstrate the role of quantum fluctuations involving remote bands, as we will show in this work, before and across the Mott-Wigner transition. 

The magnetic order of the generalized Wigner crystal phases, as well as the nature of their bandwidth and density-tuned quantum melting transitions are still a matter of debate. Previous theoretical efforts to understand moiré Wigner crystals and their evolution with interaction strength have focused on the deep crystalline regime \cite{WC_TBG_Phillips,PhillipsWigner,LiangFuMonteCarlo}, where classical Monte Carlo simulations can be used to investigate the ground state charge order at different fillings. Hartree-Fock  \cite{FengchengTMD2, DasSarmaNagaokaPRL,DasSarmaTdependence} and classical Monte Carlo studies \cite{KimMattyMonteCarlo} have addressed the competition between different charge and spin orders in the crystal phase, and its transition to metal when bandwidth or density are tuned. An analysis that includes quantum fluctuations and goes beyond mean-field is needed, however, since mean-field theory approximations  
are known to favor ferromagnetic groundstates and to overestimate the stability of 
insulating states in the proximity of metal-insulator transitions.

In this work, we report on a finite-system exact diagonalization study of semiconductor moiré materials at fractional filling factors. 
Starting from the continuum model description \cite{FengchengHubbard}, we add relevant electron-electron interactions projected to the topmost moiré band and obtain the many-body spectrum. We find,
as already suggested by experiment, that a rich set of fractional band fillings $\nu$ support correlated insulating states with tunable magnetic properties. We show that there is an overriding competition between antiferromagnetism and ferromagnetism particularly at $\nu=2/3$ band filling, which we explain using a low-energy effective spin model description and relate to a recent experiment \cite{FrustratedWignerMott}. We also address the Mott-Wigner transitions at fractional fillings, finding that as in the $\nu=1$ case they are not strongly first order. 

\section{Moiré Material Model}
\subsection{Continuum model}
\begin{figure}[h]
\centering
\includegraphics[width=0.9\linewidth]{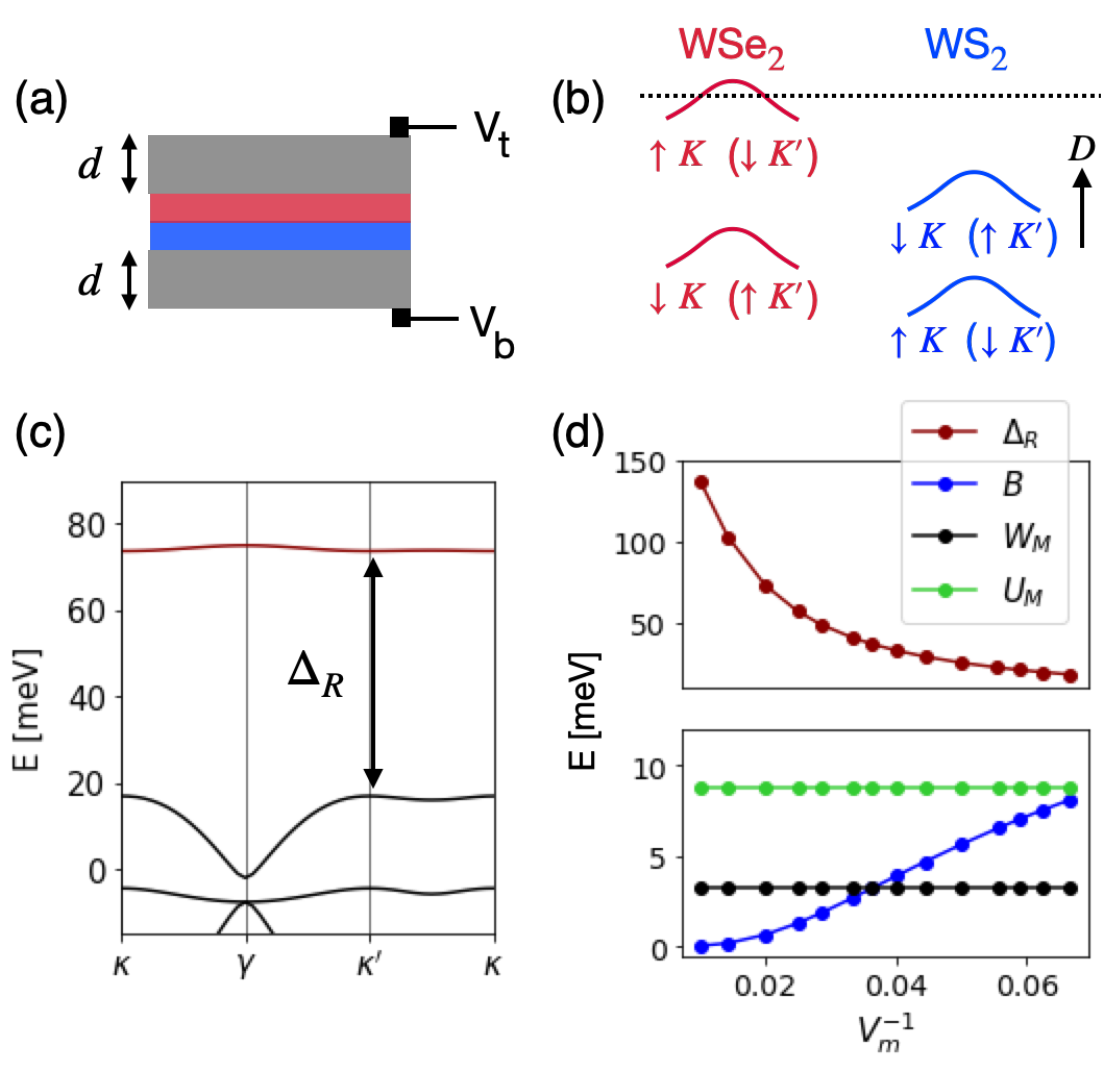}
\caption{(a) Schematics of a TMD heterobilayer with metallic gates at distance $d$ from the sample. By varying gate volt\nobreak ages a displacement field is applied. (b) Schematics of the heterobilayer in momentum space. Charge carriers populate the active WSe$_2$ band while
the presence of a WS$_2$ layer generates the moiré potential, Eq. \eqref{Moirepotential}, whose strength is modified by the displacement field $D$. (c) Example of WSe$_2$/WS$_2$ moiré minibands obtained from Eq. \eqref{Continuummodel}, with $V_m=40$ meV. (d) Evolution of the energy scales of the problem as $V_m$ is varied: the interaction strength $U_M$ (green), the kinetic energy scale $W_M$ (black), the bandwidth $B$ (blue) and the gap to the first remote $\Delta_R$ (brown).}
\label{fig:SingleParticle}
\end{figure}

Our starting point is the continuum model description of TMD heterobilayers with type-I or type-II band alignment \cite{FengchengHubbard}. In 
this case the topmost moiré band is concentrated in one of the layers, which we refer to as the active layer, depicted in red in Fig. \ref{fig:SingleParticle}(a)-(b).  
The influence of the second layer, shown in blue in Fig. \ref{fig:SingleParticle}(a)-(b), is responsible for
a moiré potential that affects charge carriers in the active layer. The moir\'e pattern can be induced by a small 
twist angle $\theta$ or a lattice mismatch $\delta$ between the two layers.  The moir\'e lattice constant
is given by $a_M = a_0/(\theta^2+\delta^2)^{1/2}$, with $a_0$ the lattice constant of the active TMD layer. To date most 
heterobilayer experiments have focused on unrotated WSe$_2$/WS$_2$ 
with a moir\'e lattice constant of $a_M\approx 8.2$ nm, or MoSe$_2$/WS$_2$, with a moiré 
lattice constant of $a_M\approx 7.5$ nm.  
Because the moir\'e lattice constant reaches a maximum, 
the system is expected to be less sensitive to twist angle disorder at zero twist angle.  

Valley and spin are locked in TMD heterobilayers and the valley (or spin)  
projected continuum Hamiltonian is given by \cite{FengchengHubbard}
\begin{align}
    \label{Continuummodel}
    H_0&=-\frac{\hbar^2}{2m^*}{\bm k}^2 +\Delta({\bm r}),\\
    \label{Moirepotential}
    \Delta({\bm r})=&2V_m\sum_{j=1,3,5}\cos({\bm b}_j\cdot{\bm r}+\psi).
\end{align}
The first term in Eq.~\eqref{Continuummodel}, which corresponds to the kinetic energy of carriers in the top moiré band, is 
diagonal in momentum space and the second term, the moiré potential, is diagonal in coordinate space.
The moir\'e potential depends on only two parameters $(V_m,\psi)$ because \cite{FengchengHubbard} of the system's $C_3$ symmetry.  The phase $\psi$ fixes the geometry of the moiré superlattice, which we take to be triangular as it is in most of the TMD heterostructures, and the strength of the moiré potential $V_m$ can be related to the experimentally tunable displacement field. For concreteness, throughout this work we take the effective mass $m^*=0.45\,m_0$ and $\psi=45^{\circ}$, corresponding to WSe$_2$/WS$_2$ \cite{LiangFuQuantumChemistry}. We take the modulation potential strength $V_m$ as an experimentally controllable parameter since it has been demonstrated to be 
sensitive to the displacement field $D$.  The form we have used for the moiré modulation parameter assumes that it varies smoothly with position on the moir\'e scale; higher harmonics in the plane-wave expansion are more important in longer period moirés in which 
relation relative to rigidly twisted bilayers is stronger.

An example of the bandstructure obtained from diagonalizing Eq. \eqref{Continuummodel} is shown in Fig. \ref{fig:SingleParticle}(c). We label the band energies of Eq. \eqref{Continuummodel} as $\epsilon_{{\bm k}}^n$, while the eigenstates can be written in a plane wave expansion as $\ket{n,\psi_{\bm k}}=\sum_{{\bm G}}z^n_{{\bm k},{\bm G}}\ket{{\bm k}+{\bm G}}$, where $n$ is a band index and ${\bm G}$ are reciprocal lattice vectors. In order to study the emerging many-body phases we consider the interacting Hamiltonian projected to the topmost moiré valence band (hence we omit band index)
\begin{align}
\label{ManyBodyHamiltonian}
    H&=\sum_{{\bm k}}\epsilon_{{\bm k}}c^{\dagger}_{{\bm k}\sigma}c_{{\bm k}\sigma}+\sum_{\substack{i,j,k,l\\
    \sigma,\sigma^{\prime}}}\frac{V_{i,j,k,l}^{\sigma,\sigma^{\prime}}}{2}\,c^{\dagger}_{{\bm k}_i\sigma}c^{\dagger}_{{\bm k}_j\sigma^{\prime}} c_{{\bm k}_l\sigma^{\prime}}c_{{\bm k}_k\sigma},
\end{align}
where the Coulomb matrix elements are given by 
\begin{align}
\label{InteractionME}
    V_{i,j,k,l}^{\sigma,\sigma^{\prime}}=\frac{1}{A}\sum_{\substack{{\bm G}_i,{\bm G}_j\\{\bm G}_k,{\bm G}_l}}\left(z^{*}_{{\bm k}_i,{\bm G}_i} z^{*}_{{\bm k}_j,{\bm G}_j}z_{{\bm k}_k,{\bm G}_k}z_{{\bm k}_l,{\bm G}_l}\right) \frac{2\pi e^2}{\varepsilon\,  q}.
\end{align}
The summation involving the term in brackets results from the projection of interactions to a single band and can be rewritten in terms of the {\it form factors}, as described elsewhere \cite{Repellin_FormFactors}. In Eq.~\eqref{InteractionME}, $ q = |{\bm k}_i+{\bm G}_i - {\bm k}_k-\bm{G}_k| $ is 
the momentum transfer, $A$ is the area of the system, $\varepsilon$ the effective dielectric constant,
and total momentum conservation is implicit. 

\subsection{Exact diagonalization methodology}
In this article we take the same approach as in previous work \cite{MoireMIT,Nonlocal}, utilizing exact diagonalization (ED) to solve the many-body Hamiltonian. We present our results in phase diagrams that depend on dimensionless parameters obtained by taking ratios of the relevant energy scales of the system. In particular, three energy scales can be identified; the moiré potential depth $V_m$, the interaction strength $U_M=e^2/(\varepsilon a_M)$, and the kinetic energy scale $W_M=\hbar^2/m^*a_M^2$. 
By varying the two ratios of these three scales, we can simulate any heterobilayer
as long as the moiré period is much larger than the microscopic lattice constant.
This ensures that our conclusions apply for arbitrary heterobilayers as long as their low energy physics is captured by the continuum model, Eq. \eqref{Continuummodel} and 
that the shape of the moir\'e potential is similar to that of a triangular lattice model.
The energy gap to the remote moiré bands $\Delta_R$ can, in principle, be viewed as another relevant energy scale. We take parameter values such that this scale is always larger than all other scales involved, as can be seen in Fig. \ref{fig:SingleParticle}(d), justifying the single band projection of Eq. \eqref{ManyBodyHamiltonian}. 
As we have pointed out above, however, interactions renormalize bands more at higher electron densities.  For this reason the single-band 
approximation should be treated with caution for fillings $\nu>1$,
where remote band mixing is often relevant.

The model we study has orbital and spin degrees of freedom, discrete triangular lattice translational symmetry, $SU(2)$ spin-rotational invariance, and no spin-orbit coupling.
The Hilbert space can be divided into smaller subspaces with total momentum ${\bm K}$ with discrete values determined by the number of moiré unit cells $M$, total spin ${\bm S}$, and azimuthal spin $S^z$. The basis is constructed in an occupation number representation, distributing particles among single-particle states
labeled by $S^z$ quantum number and quasi-particle $(k_x,k_y)$ momentum. 
The total number of possible configurations $N_{conf}$ for particles distributed on $M$ single particle states with $N_\uparrow$ spins up and $N_\downarrow$ spins down is determined by a product of binomial coefficients,  $N_{conf}=\binom{M}{N_{\uparrow}} \cdot \binom{M}{N_{\uparrow}} $. 
The many-body Hamiltonian projected to a given total momentum ${\bm K}$ is diagonalized in 
$S^z$ subspaces. We do not rotate the Hamiltonian matrix to a ${\bm S}$ basis as this is an additional computational cost, and instead determine the ground state total total spin $S$ by identifying 
multiplets from the $S^z$-dependent energy eigenvalues.  The total spin $S$ assignments 
have been confirmed by calculating the spin structure factor $\mathcal{S}({\bm q}=0)$. For a given momentum, the $S$-multiplet structure implies
that the largest subspace corresponds to the lowest possible $S^z$ which contains states
with all values of total spin ${\bf S}$.

All of our exact diagonalization calculations are limited only by the maximal matrix size of a given subspace, with the largest subspace corresponding to the lowest possible $S^z$. The results presented here correspond to systems containing $M=9,12$ and $16$ moiré unit cells. Despite the limited system sizes that can be reached with the exact diagonalization method, important information can still be extracted concerning the behaviors of charge gaps and the nature of the magnetic order of insulating states. 
Numerical non-perturbative approaches, 
like the one taken in this paper or DMRG, 
are particularly important in the quantum melting regime, where Hartree-Fock is known to overestimate the stability of the insulating phase. 

\section{Finite-Size Phase Diagrams}
\begin{figure}[h]
\centering
\includegraphics[width=\linewidth]{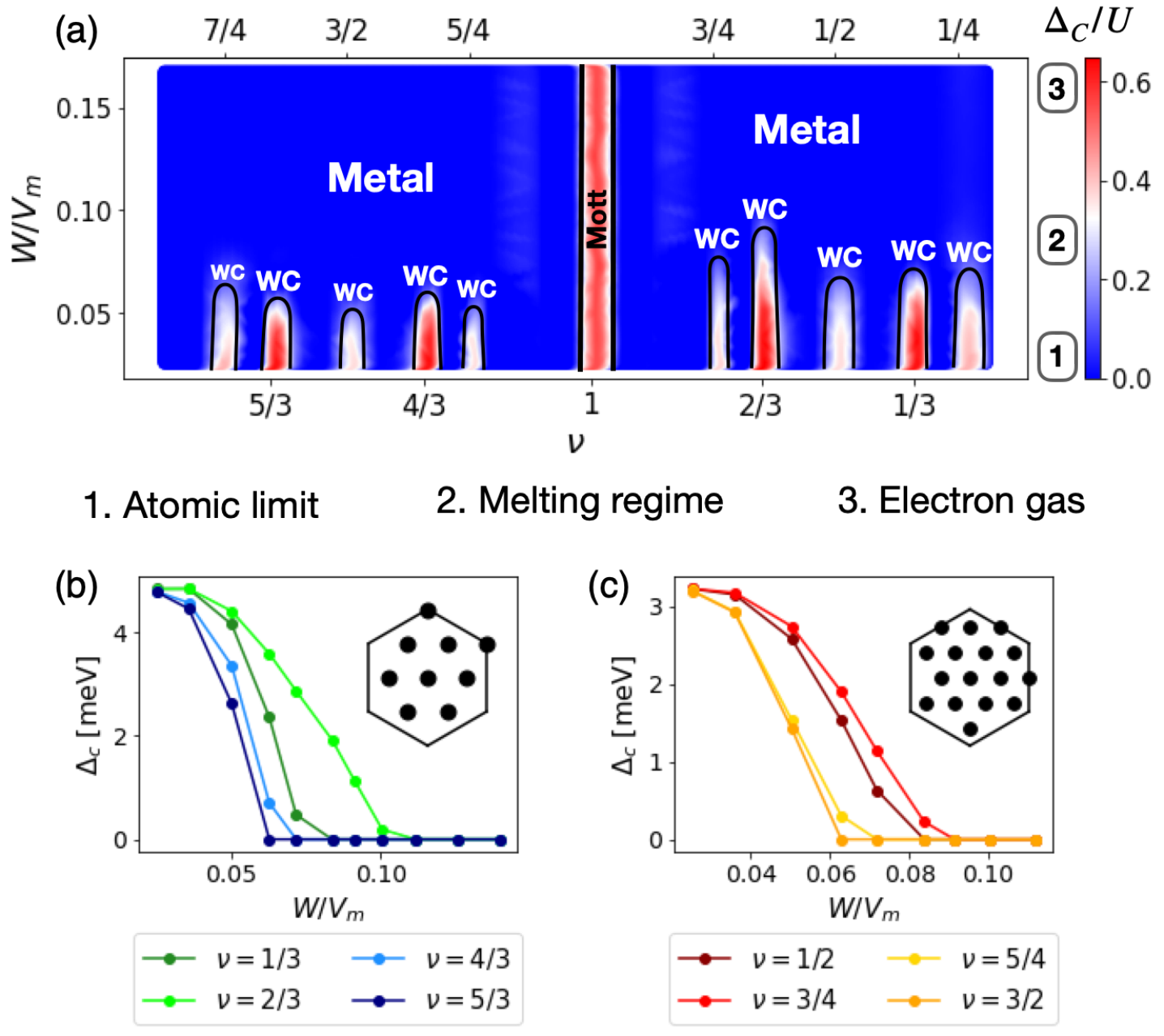}
\caption{(a) Exact-diagonalization phase diagram for TMD heterobilayers as a function of filling factor $\nu$ and $W/V_m$ (kinetic energy to moir\'e potential depth ratio). We perform calculations on finite-size systems with $N$ electrons in $M$ unit-cells. Results shown are obtained at  $M=9$ for fillings multiples of $\nu=1/3$ and $M=16$ for fillings multiples of $\nu=1/4$. Incompressible (gapped) states are 
clearly present at several values of $\nu=N/M$ specified in the main text, with the color scale indicating the magnitude of the
charge gap $\Delta_c$ relative to the interaction scale $U$, at $\varepsilon=20$. Our results are for a discrete set of $(N,M)$ and we do an interpolation to obtain a continuous plot, which is saturated for clarity. (b)-(c) Evolution of the 
charge gaps for selected values of $\nu$. The gaps coincide with known classical values 
in the atomic limit and vanish at various different values of $W/V_m$ in the melting regime. Insets show the finite geometries used.}
\label{fig:PhaseDiagram}
\end{figure}

A phase diagram for TMD heterobilayers as a function of filling factor and the kinetic-energy-scale to moiré-depth ratio $W/V_m$ is shown in Fig. \ref{fig:PhaseDiagram}(a) . The color scale specifies the ratio of the charge gap to the
Coulomb interaction-energy scale $\Delta_c/U$. 
A finite value of the charge gap indicates an incompressible state, while a vanishing charge gap indicates a metallic state. The charge gap is extracted from the many-body spectrum via the relation $\Delta_c(N)=E(N+1)+E(N-1)-2E(N)$, where $N$ is the number of particles in the system and $E$ is the energy of a many-body ground state obtained from diagonalization of the Hamiltonian in Eq. \eqref{ManyBodyHamiltonian} for a system with a finite number of moiré unit cells $M$. 
When interactions are sufficiently strong, we verify the presence of a
Mott insulating state at half-filling $\nu=N/M=1$, and also of a set of incompressible states at the fractional fillings: $\nu=1/3,2/3,4/3,5/3,1/4,1/2,3/4,5/4,3/2,7/4$ \cite{Note1}.  All insulating states 
become metallic when $V_m$ is decreased below a critical value, in qualitative agreement with experimental results. 
Fig. \ref{fig:PhaseDiagram}(b) shows the evolution of the charge gap in meV with $W/V_m$ for selected multiples of $ \nu= 1/3$ and $ \nu= 1/4$. The typical gap values in the localized limit $\Delta_c \sim 3-5$ meV, are in accord with those measured in experiment \cite{ContinuousWigner,FrustratedWignerMott}.

At each fractional filling factor we can follow the evolution of the ground state 
from the atomic limit as the bandwidth increases.  For very narrow bands, the physics reduces to that of a triangular lattice-gas model. At rational filling factors the charge distributions that minimize the interaction energy break translational symmetry and have an energy gap for unbound electron-hole pairs. As the bandwidth increases, quantum fluctuations induce interactions between electron spins that is usually expected to 
yield a ground state with magnetic order. Quantum fluctuations can potentially change the preferred charge order \cite{LiangFuMonteCarlo} from that of the atomic limit and they eventually dominate, driving a transition to a metallic state. In the following we first make a number of general remarks about our numerical results and 
then focus on the filling factors $\nu=1/3$ and $\nu=2/3$ that have the most 
prominent insulating states in experiment \cite{CornellWigner,CaliforniaWigner,Berkeley,CornellWignerStripe,STM_Wigner}.
\subsection{General Trends}
We can identify three main regimes in our phase diagram, Fig. \ref{fig:PhaseDiagram}(a)
\begin{enumerate}
    \item {\it Atomic limit} -- $W/V_m \ll 1$. This limit corresponds to a perfectly flat band with 
    $t=0$, equivalent to a classical electron gas on a lattice created by the strong moiré potential. Electrons are expected to localize on a subset of superlattice sites so as to minimize on-site and near-neighbor interactions, giving rise to generalized Wigner crystal states. Classical Monte Carlo \cite{LiangFuMonteCarlo,CaliforniaWigner} and Hartree-Fock \cite{HoneycombWCTenessee} calculations can address the specific form of the charge order, which
    depends on the filling. We see in Fig. \ref{fig:PhaseDiagram}(b) that the charge gaps at all multiples of $1/3$ and $1/4$ are equal in the atomic limit. The many-body ground state manifold contains a large number of nearly degenerate states, corresponding to different spin states on the same sublattice of occupied sites.  As the atomic limit is approached,
    the spacings between these levels become too small 
    to allow numerical determination of the ground state magnetic order. The small spacing of these levels implies that the full entropy of the spin subspace will
    be realized at a low temperature. 
    
    \item {\it Melting regime} -- $W/V_m \sim 0.05-0.1$. In the intermediate regime, the competition between electron localization due to long-range Coulomb interaction and quantum fluctuations controls the generalized Wigner crystal melting.  Quantum fluctuations of charge distributions are enabled by the hopping term, $t$.  When $t$ increases from the atomic limit,
    interactions between spins located on different sites strengthen. 
    The ground state spin manifold then broadens sufficiently to allow the magnetic properties of some crystal states to be addressed. Interestingly, depending on the dielectric constant value $\varepsilon$ and the filling fraction $\nu$, we can obtain either ferromagnetic or antiferromagnetic states.

    \item {\it Electron gas limit} -- $W/V_m \gg 1$: This is the limit of weak moiré modulation. The system will resemble a 2DEG and the ground state is determined by the electron density. The question of how Wigner crystallization in the 2DEG is modified by adding an underlying lattice as a small perturbation is not a trivial one and the properties expected for electron crystals in the 2DEG are significantly modified. We obtain vanishing charge gaps at all fillings corresponding to Fermi liquid states as can be seen in Fig. \ref{fig:PhaseDiagram}(a) (in Appendix B we show examples of occupation distributions where the Fermi surfaces can be identified). This is a limit where Hartree-Fock calculations do not give reliable results \cite{Vignale_Book}.
\end{enumerate}

In focusing on filling fractions $\nu=1/3$ and $\nu=2/3$, we will mainly discuss results for a system containing $M=9$ moiré unit cells in the main text and present some additional results for $M=12$ in Appendix B. Despite the small system size, the $M=9$ geometry captures the charge distribution observed in experiments \cite{STM_Wigner,STM_Wigner_Excitations}, hence it can give us some insight into how increasing hopping amplitudes from the localized limit initially determines the magnetic ground state and then, ultimately, drives melting.

In order to determine the nature of the magnetic ground states, translational symmetry breaking can be tested by evaluating the static spin-spin correlation function 
\begin{align}
    \xi({\bm r},{\bm r'})=\langle {\bm S} ({\bm r}) \cdot {\bm S}({\bm r'})\rangle,
\end{align}
and using it to calculate a static spin structure factor defined as 
\begin{align}
    \mathcal{S}({\bm q})= \frac{1}{A^2} 
    \int d{\bf r} \int d{\bf r'} \; e^{-i\,{\bm q} \cdot({\bf r}-{\bm r'}) } 
    \xi({\bm r},{\bm r'}).  
\end{align}
In these equations $\bm{q}$ is an extended zone wavevector,
$A$ is the sample area, and the expectation values are taken in the many-body ground state.
Finite size peaks in this quantity at wavevectors that are not reciprocal lattice vectors
indicate broken translational symmetry.  The inverse Fourier transform of $\mathcal{S}({\bm q})$,
\begin{align}
    \xi({\bm{r}})= \frac{1}{A} \sum_{\bm{q}} \; e^{i\,{\bm q} \cdot {\bf r}}\, \mathcal{S}({\bm q}) 
    =\frac{1}{A} \int d{\bm{x}} \; \xi({\bm x},{\bm x}+{\bm r}),  
\end{align}
is used below to characterize the correlations between spins at positions separated by ${\bm{r}}$.

Additional insight is provided by the number of finite-size many-body 
eigenvalues in different energy ranges. The Hilbert space can be divided into a singly-occupied subspace (low energy sector) and a doubly-occupied subspace (high energy sector) 
analogous to the lower and upper Hubbard bands for half-filling. Since the spectrum connects 
adiabatically to the atomic limit, the number of states with energy smaller than $\sim U_0$
is equal to the dimension of the full single-occupancy Hilbert space. For filling $\nu=N/M$, the subspace of the single-occupancy Hilbert space with $N_{\uparrow}$ spins up and $N_{\downarrow}$ spins down has dimension given by 
\begin{align}      d(N_{\uparrow},N_{\downarrow})=\frac{M!}{N_{\uparrow}!\,N_{\downarrow}!(M-N)!}.
\end{align}
By considering all possible configurations $(N_{\uparrow},N_{\downarrow})$ we obtain the total 
dimension of the single occupancy, low-energy, sector.
\subsection{$\nu=1/3$ Band Filling}
\begin{figure}[h!]
\centering
\includegraphics[width=\linewidth]{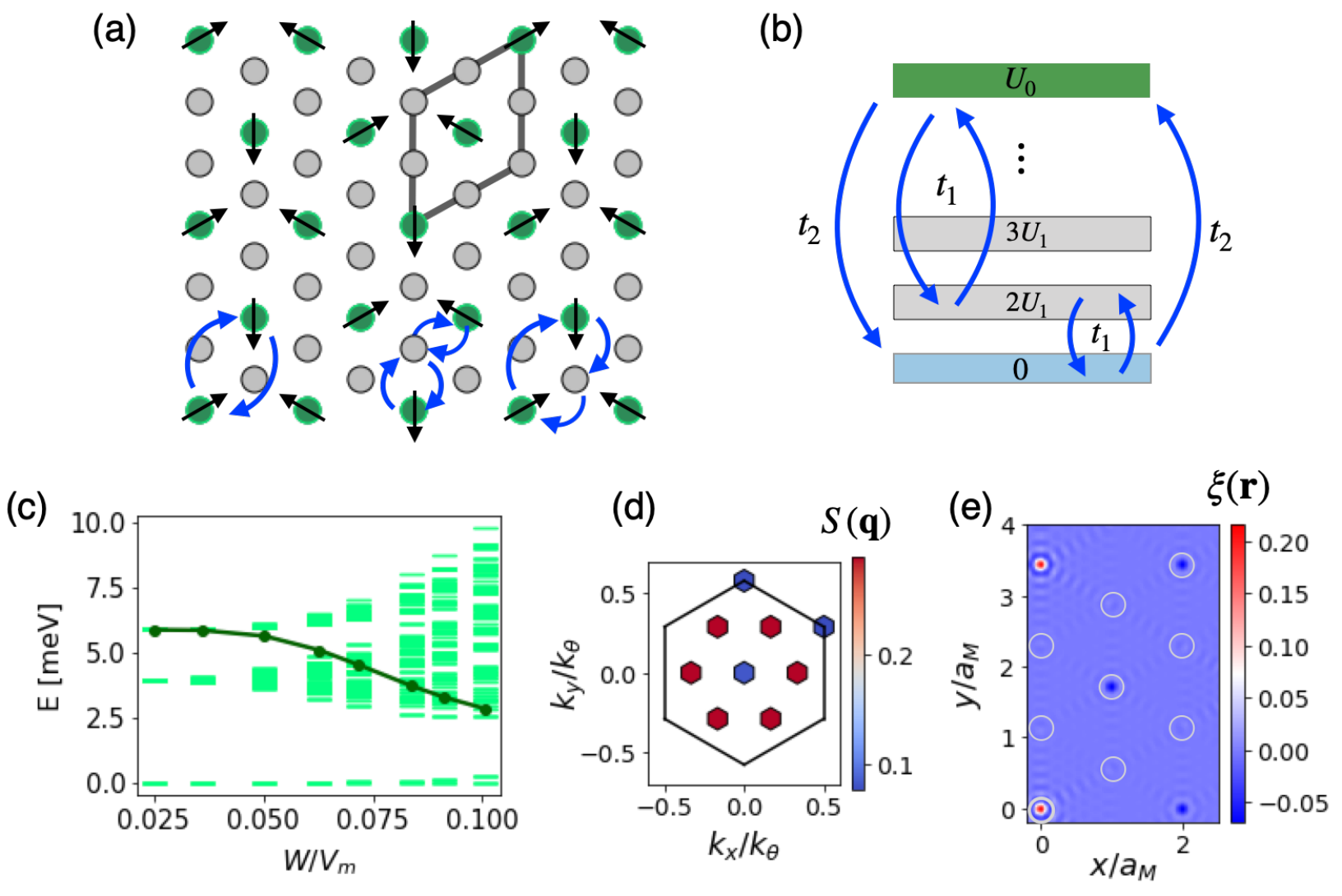}
\caption{Generalized Wigner crystal state at $\nu=1/3$. (a) Real space configuration indicating charge order (green) and localized spins forming a $120^{\circ}$-Néel state. 
The magnetic unit cell is indicated in gray. Super-exchange processes of order $t_2^2$, $t_1^4$ and $t_1^2\,t_2$ are represented by blue arrows. (b) Schematics of the low-energy spectrum and the double-occupancy manifold with energy $\sim U_0$, with the hopping processes connecting different sectors. (c) Many-body low-energy spectrum as a function of $W/V_m$ showing the evolution of the three bands corresponding to the ground state manifold, bound particle-hole pairs (excitons) and itinerant charged excitations. 
The charge gap evolution is shown as a green solid line (We have added a Coulomb-blockade correction that brings $\Delta_c$ and the bottom of the charged excitation band 
of many-body excitations into coincidence). (d) Structure factor and (e) spin-spin correlation function calculated at $V_m=30$ meV. Lattice sites in (e) are indicated as white circles.}
\label{fig:Onethird}
\end{figure}

In Fig. \ref{fig:Onethird}(a) we illustrate
the spatial configuration of the ground state for $\nu=1/3$. In the atomic limit ($t=0$) all spin configurations are degenerate and the ground state 
charge distribution corresponds to a triangular lattice with periodicity $\sqrt{3}\,a_M$ (represented as green sites).  This is the only charge configuration that avoids the 
$U_1$ nearest neighbor interaction scale.  This charge order has been measured by STM \cite{STM_Wigner,STM_Wigner_Excitations}. Other theoretical studies \cite{FengchengTMD2,DasSarmaNagaokaPRL,DasSarmaTdependence} 
have also found the charge configuration of Fig. \ref{fig:Onethird}(a) in the highly localized limit.

Starting from the atomic limit, there is a regime of finite $t$ in which the charge order pattern and the insulating state gap survive, but quantum fluctuations induce interactions between localized spins that 
lift the large ground state spin degeneracy. 
From our ED results we obtain a ground state with minimum total spin, indicating an antiferromagnetic state. The magnetic configuration in this regime is also shown in Fig. \ref{fig:Onethird}(a). We confirm that spins form a $120^{\circ}$-Néel antiferromagnetic state by evaluating the structure factor $\mathcal{S({\bf q})}$ and its inverse Fourier transform $\xi({\bm r})$, presented in Fig. \ref{fig:Onethird}(d) and (e) respectively. We observe structure factor peaks at the middle inner points ${\bm \vartheta}$ of the Brillouin zone, indicating a magnetic unit cell with 9 sites. From the correlation function $\xi({\bm r})$ we see that all first neighbors of an occupied site are empty, while the spin orientations on the second neighboring sites form an angle larger than $\pi/2$ with respect to the occupied site, which translates into negative values. 
In the limit of classical spins, for this triangular Néel state, evaluation of the structure factor at high symmetry points results in a finite value only at ${\bm \vartheta}$
\begin{align}
    \mathcal{S}({\bm \gamma})=\mathcal{S}({\bm \kappa})=0, ~~~~ \mathcal{S}({\bm \vartheta})=\frac{1}{72}.
\end{align}
When quantum fluctuations are not too strong, these classical estimates are still approximately valid \cite{Dagotto}. Therefore the appearance of peaks at ${\bm \vartheta}$ in Fig. \ref{fig:Onethird}(d) is in agreement with the 
indicated spin configuration.

The many-body spectrum separates into a high-energy sector associated with double occupations and characteristic energy $U_0$ and a low-energy sector of configurations with no double occupations. For $N=3$ particles on $M=9$ sites, the full  Hilbert space has dimension 816 and the low-energy sector of single-occupancy consists of 672 states. Fig. \ref{fig:Onethird}(c) shows the many-body energies as a function of $W/V_m$, obtained by ED, corresponding to the low-energy single-occupancy Hilbert space, that in turn separates into three bands. 

The lowest band, or ground state manifold, contains 24 states that are degenerate in the atomic limit. This number can be understood by noting that there are three inequivalent Wigner crystal configurations at $\nu=1/3$--filling, corresponding to choosing one of the three sites to occupy within the unit cell. 
For each of these states there is a multiplicity of $2^3$ when the spin degree of freedom is taken into account. As $t$ is increased (right limit of the plot) the degeneracy is lifted. In contrast to the half-filled case, the $\nu=1/3$ state has a branch of excitonic states with energies that lie below the charge gap, indicated by a dark green line with dots, as can be observed in our spectrum. The number of these particle-hole excitations is $N_{ex}=432=24\times3\times6$, corresponding to the product of the number of states in the
ground state manifold, the number of electrons, and the number of neighboring empty sites around each filled site. Finally, the third band in Fig. \ref{fig:Onethird}(c) is the manifold of itinerant charged excitations.  Its minimum coincides with the charge gap $\Delta_c$ in the atomic limit. This band has 216 states for the system size considered, corresponding to configurations where the three particles occupy neighboring sites.
The multiplicity of the itinerant charged excitation branch grows most quickly as the system size is increased.

From the mapping of the TMD continuum model to an extended triangular Hubbard model \cite{FengchengHubbard,Nonlocal}, we have derived an effective spin model that describes the magnetic ground state we observe. The Heisenberg interaction between localized spins in the configuration shown in Fig. \ref{fig:Onethird}(a), up to order $t_1^4/U_0U_1^2$, is 
\begin{align}
    \label{J_Onethird}
    J_1(1/3)=\frac{4t_{\text{eff}}^2}{U_0}-\frac{2t_1^2t_2}{U_1^2}+\frac{4t_1^4}{3U_1^3}-2X_2,
\end{align}
where $t_1$ and $t_2$ stand for the first and second nearest-neighbor hopping
parameters, $X_2$ is the second-neighbor direct exchange and we have defined an effective second-neighbor hopping $t_{\text{eff}}=t_2-t_1^2/U_1$. Fig. \ref{fig:Onethird}(a)-(b) show a schematic of the 
virtual state processes that contribute to $t_{\text{eff}}$. This virtual second-neighbor hop process
does give a contribution to the effective spin model, Eq. \eqref{J_Onethird}, that 
is similar to the analogue half-filling $t_1$ process. 

Hopping of an arbitrary particle to a neighboring site by $t_1$, increases the energy by $2U_1$ (we neglect the long range part of Coulomb interaction in this analysis, for simplicity). Note that terms of order $t_1^2$ do not influence spin interactions, but repeated $t_1$ and $t_2$ hops without double-occupying a site yield the second and third terms in Eq. \eqref{J_Onethird}. The value of $J_1(1/3)$ as a function of interaction strength is plotted in Fig. \ref{fig:Magnetism}(a) below. 

\subsection{$\nu=2/3$ Band Filling}
\begin{figure}[h!]
\centering
\includegraphics[width=\linewidth]{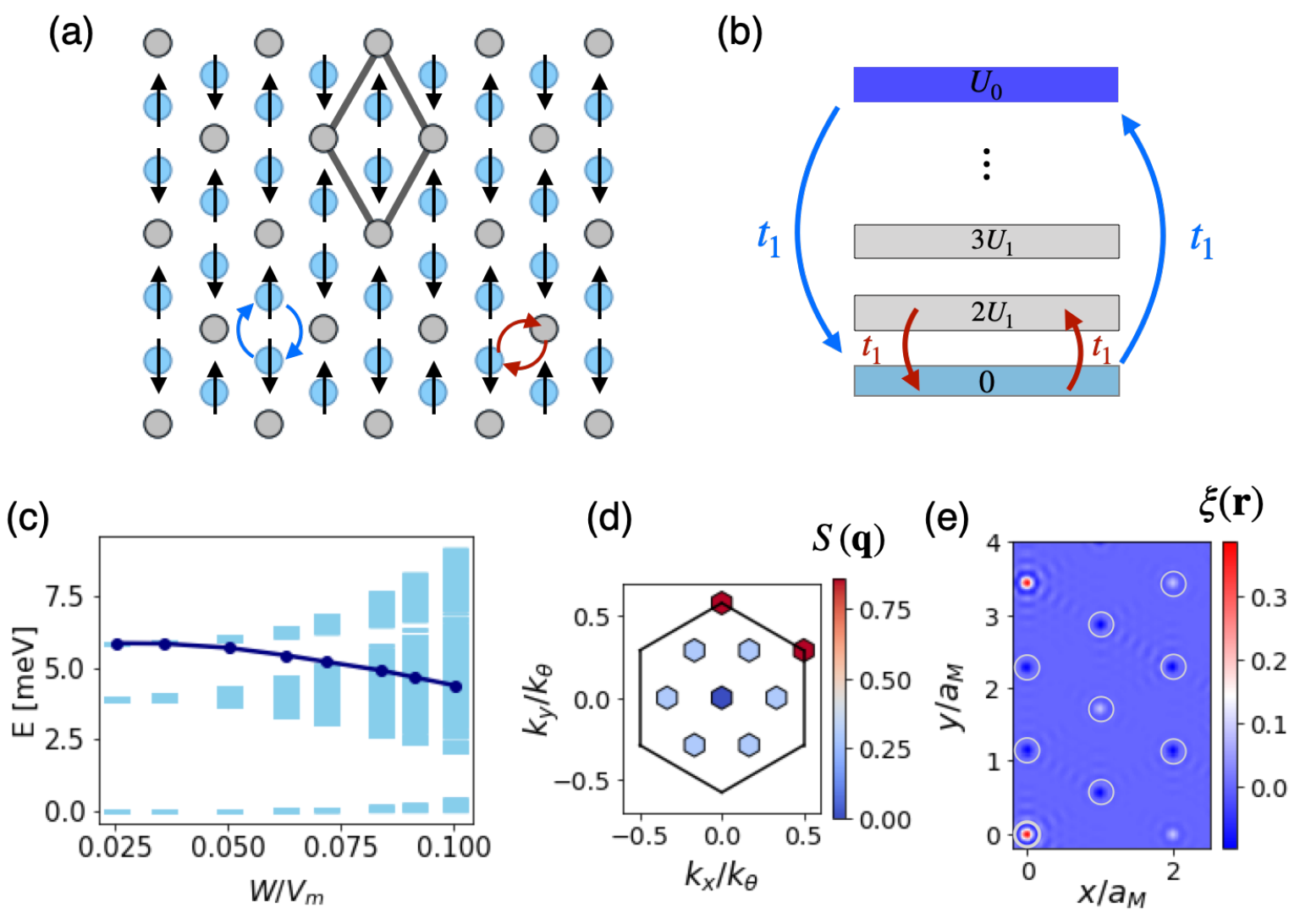}
\caption{Generalized Wigner crystal state at $\nu=2/3$. (a) Real space configuration, which effectively forms a honeycomb lattice (blue sites) with near-neighboring occupied sites having opposite spins. Processes contributing to the spin coupling and to the formation of the excitonic band are depicted in blue and red respectively. (b) Schematics of the three low-energy bands and the double-occupancy manifold, the dominating super-exchange process of order $t_1^2$ is represented by the blue arrows, however direct exchange $X_1$ also gives a significant contribution to the spin coupling. (c) Many-body low-energy spectrum as a function of $W/V_m$ showing three bands: the ground state manifold, particle-hole pair excitations and charged excitations. The charge gap evolution is shown as a blue solid line. (d) Structure factor and (e) Spin-spin correlation function, calculated at $V_m=30$ meV.}
\label{fig:Twothirds}
\end{figure}

The spatial configuration of the ground state for $N=6$ particles on $M=9$ sites (for the same parameters as those of  Fig. \ref{fig:Onethird}) is shown in Fig. \ref{fig:Twothirds}(a). In the atomic limit the charge distribution forms a honeycomb lattice, in agreement with recent STM measurements \cite{STM_Wigner}. When quantum fluctuations are turned on, the ground state has antiferromagnetic order,
consistent with the measured Curie-Weiss (CW) temperature 
indicator \cite{FrustratedWignerMott} on a similar heterobilayer system. 

The structure factor, shown in \ref{fig:Twothirds}(d), has peaks at ${\bm \kappa}$, indicating a magnetic unit cell with three sites, while also showing smaller non-zero values for the middle points ${\bm \vartheta}$. The function $\xi({\bm r})$ is illustrated in Fig. \ref{fig:Twothirds}(e). The second neighbor sites have a positive value indicating same spin orientation, while the six nearest neighbors of an occupied site have negative values. We interpret these small negative values as the average of three occupied sites with opposite spin orientation and three empty sites. These spin correlation function results seem to indicate that at $\nu=2/3$--filling, there is greater virtual occupation of empty sites compared with the $\nu=1/3$--filling case. A classical analysis of the structure factor for the spin configuration in Fig.  \ref{fig:Twothirds}(a) yields the following values for high-symmetry points
\begin{align}
    \mathcal{S}({\bm \gamma})=\mathcal{S}({\bm \vartheta})=0, ~~~~ \mathcal{S}({\bm \kappa})=\frac{1}{12},
\end{align}
allowing us to conclude in favor of the state in Fig.  \ref{fig:Twothirds}(a) when quantum fluctuations are not strong. 

For $N=6$ particles on $M=9$ sites, the full Hilbert space has dimension 18564, while the dimension of the single-occupancy Hilbert space is 5376. Fig. \ref{fig:Twothirds}(c) shows the low-energy many-body spectrum, which separates into three bands
as in the $\nu=1/3$ case, with a higher-energy 
charged sector, and a mid-energy region containing excitonic states with 
bound electron-hole pairs, indicated by red arrows in Fig. \ref{fig:Twothirds}(a)-(b), and the ground state manifold. The ground state band has 192 states, corresponding to three inequivalent Wigner crystal configurations (the conjugate configurations of the $\nu=1/3$ case) and a multiplicity of $2^6$ for each, when the spin degree of freedom is accounted for. The electron-hole bound pair sector has $N_{ex}=3456=192\times6\times3$ states, which result from the product of the number of states in the ground state manifold, the number of electrons and the number of empty sites where each electron can hop. The higher band contains the charged excitations and has 1728 states, which correspond to the conjugates of the charge distributions forming the higher band in the $\nu=1/3$ case, times the spin multiplicity $2^6$. In this case the Heisenberg coupling constant of the effective spin honeycomb model for the configuration in Fig. \ref{fig:Twothirds}(a), at lowest order, is
\begin{align}
\label{J_twothirds}
    J_1(2/3)=\frac{4t_1^2}{U_0-U_1}-2X_1,
\end{align}
where $X_1$ is nearest neighbor direct exchange. Virtual hopping to a double occupied site by $t_1$ is the main process responsible for antiferromagnetism, the first term in Eq. \eqref{J_twothirds}, shown by blue arrows in the cartoon of Fig. \ref{fig:Twothirds}(a)-(b). The value of $J_1(2/3)$ as a function of interaction strength is plotted in Fig. \ref{fig:Magnetism}(b).

\subsection{Tuning magnetic properties}

\begin{figure}[h!]
\centering
\includegraphics[width=\linewidth]{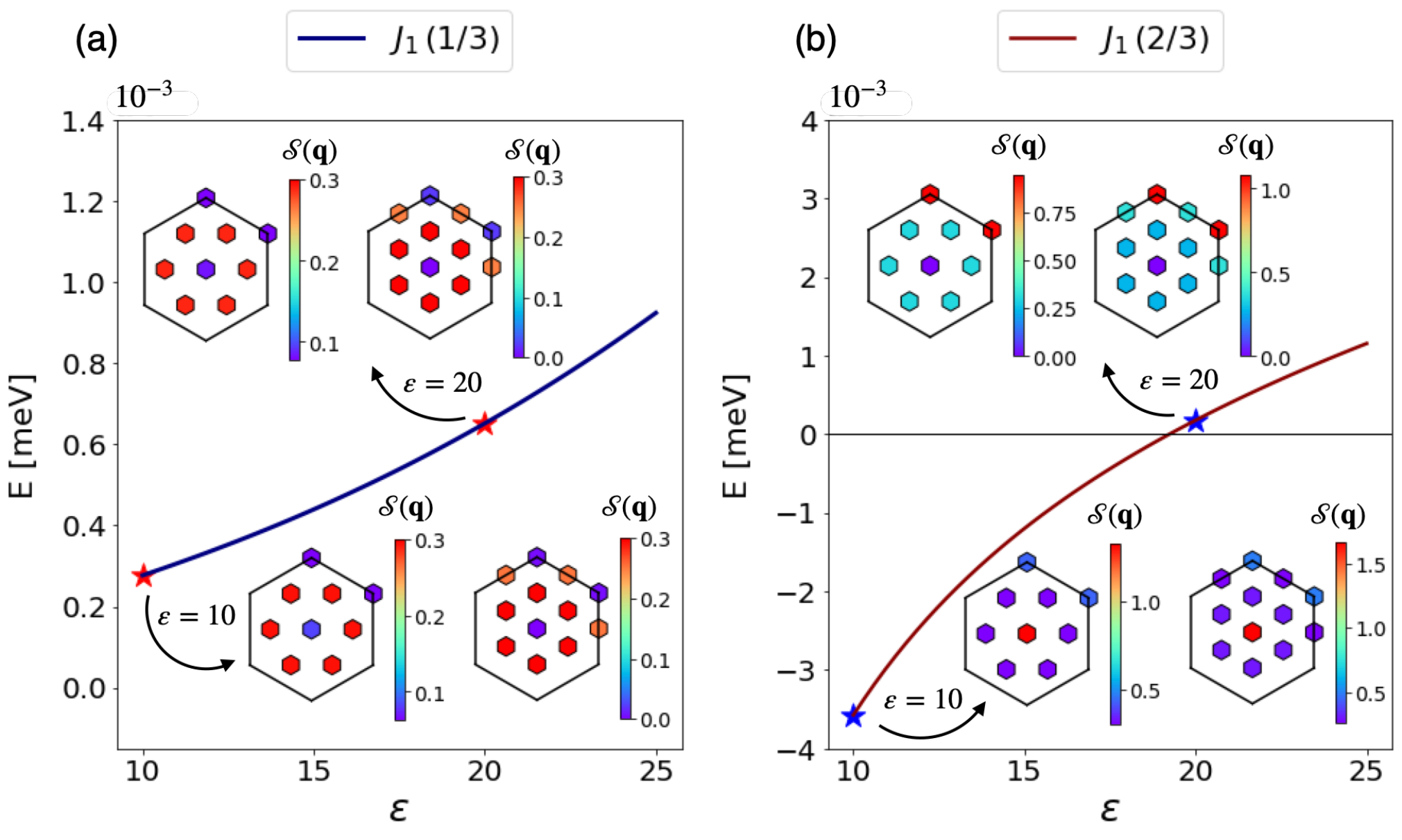}
\caption{Competing magnetic interactions in generalized Wigner crystals. Solid lines show the effective spin couplings, (a) $J_1(1/3)$ and (b) $J_1(2/3)$, as a function of $\varepsilon$. Insets correspond to structure factors for system sizes $M=9,12$ and two values of dielectric constant $\varepsilon=10,\,20$. Peaks in (a) always remain at ${\bm \vartheta}$, consistent with a magnetic unit cell of 9 sites with antiferromagnetic order. In contrast, at $\nu=2/3$-filling the main structure factor peak changes location from ${\bm \kappa}$ at $\varepsilon=20$ to ${\bm \gamma}$ at $\varepsilon=10$, indicating a transition from a an antiferromagnetic to a ferromagnetic state. Results shown in (a) and (b) have been obtained for $V_m=30$ meV and $V_m=40$ meV respectively.}
\label{fig:Magnetism}
\end{figure}

The magnetic honeycomb pattern found at $\nu=2/3$, shown in Fig. \ref{fig:Twothirds}(a), is sensitive to model parameters. If the value of the dielectric constant $\varepsilon$ is increased, the ground state transits from antiferromagnetic to ferromagnetic.  The 
dielectric function increase can be accomplished experimentally by modifying the substrate or varying the distance from active device to electrical gates.
This change of properties is illustrated
in Fig. \ref{fig:Magnetism}, where we present structure factors for two values of $\varepsilon$ for both fillings $\nu=1/3$ and $2/3$. (We also include results corresponding to $M=12$.)  For $\nu=1/3$, shown in Fig. \ref{fig:Magnetism}(a), the main peaks are at the middle points, ${\bm \vartheta}$, for the two system sizes and for both values of $\varepsilon$, in agreement with the state of Fig. \ref{fig:Onethird}(a). On the other hand, structure factors for $\nu=2/3$ change qualitatively from $\varepsilon=20$ to $\varepsilon=10$, as seen in Fig. \ref{fig:Magnetism}(b). In particular, we observe peaks at ${\bm \gamma}$ in the structure factors for $\varepsilon=10$ in both system sizes, characteristic of a ferromagnetic state. Smaller peaks at ${\bm \kappa}$ can also be observed, indicating that the system still breaks the moiré lattice translation symmetry. For $\varepsilon=20$, the main peaks in the structure factor are at ${\bm \kappa}$, in agreement with the state depicted in Fig. \ref{fig:Twothirds}(a).

The behavior at $\nu=2/3$ can be qualitatively understood in terms of the effective spin model of Eq. \eqref{J_twothirds}. As the value of $\varepsilon$ is decreased, the contribution from direct exchange starts to dominate, eventually turning the coupling constant negative and driving the ground state ferromagnetic. We show in Fig. \ref{fig:Magnetism}(b) that $J_1(\nu=2/3)$ changes sign as $\varepsilon$ decreases. For the other filling, $\nu=1/3$, the effective spin coupling $J_1(\nu=1/3)$ remains positive for all values of $\varepsilon$ considered, as seen in Fig. \ref{fig:Magnetism}(a). However, because the main processes determining the magnetic properties have a much smaller energy scale for $J_1(\nu=1/3)$, it is almost one order of magnitude smaller than  $J_1(\nu=2/3)$ at the same $V_m$. This trend agrees with a recent experiment by the Cornell group \cite{FrustratedWignerMott}, where the Curie-Weiss temperature measurements for $\nu=2/3$--filling clearly confirmed antiferromagnetic order, while for $\nu=1/3$--filling the result is inconclusive due to the very small values obtained. The experiment also found that the Curie-Weiss temperature
{\it vs.} $\nu$ has a strong local minimum at $\nu=2/3$, suggesting  a competition between super-exchange and direct exchange. Our simple finite size calculations allow us to reproduce this phenomenology. We note that in \cite{FrustratedWignerMott} a similar effective honeycomb spin model with positive first-neighbor coupling but negative second-neighbor coupling was proposed, and it can also give rise to a ferromagnetic groundstate.

\section{Discussion}

Generalized Wigner crystal states are ubiquitous in semiconductor moiré materials at fillings $\nu\neq 1$, indicating that extended-range Coulomb interactions play a more relevant role in those systems than has been recognized in atomic materials,
which are often successfully described by the standard Hubbard model. The crystal states can give rise to rich physics due to an interplay between spin and charge order. We have numerically explored how the effects of weak quantum fluctuations affect the charge order observed experimentally in the localized limit at fillings $\nu=1/3$ and $\nu=2/3$ of triangular moiré superlattices. In particular, we found a tunable magnetic ground state at $\nu=2/3$--filling, which is more sensitive to the superexchange-exchange competition of localized spins than its $\nu=1/3$ counterpart. This delicate competition,
that is also suggested by experiment, indicates that it would be possible to investigate the antiferromagnet-to-ferromagnet transition that is expected at $\nu=1$ \cite{Nonlocal} also at $\nu=2/3$. 

By further increasing the band dispersion, we addressed the melting of the Wigner crystal states. The nature of the Mott-Wigner transition \cite{ContinuousWignerDebanjan,ContinuousWigner} between insulating broken translations symmetry states and metallic states with no broken symmetries is a fundamentally important issue.  
Musser {\it et al.} \cite{ContinuousWignerDebanjan} theoretically explored the possibility of 
two continuous transitions with an intermediate spin liquid phase. 
Recent experiments, which are sensitive mainly to the charge gaps, have found that the transitions 
appear to be continuous as the moir\'e displacement field is varied \cite{ContinuousWigner}. From our results, Fig. \ref{fig:PhaseDiagram}, we can only 
conclude that the Mott--Wigner transition is not strongly first order, in the sense 
that the jump in the charge gap upon melting is very small compared to the atomic limit gap. 

The one-band models we study have particle-hole symmetry in the $t=0$ atomic limit which guarantees identical charge gaps at filling factors $\nu$ and $2-\nu$. Hopping on a triangular lattice at finite $t$ violates this symmetry as we see for instance in Fig. \ref{fig:PhaseDiagram}(b)-(c), which show larger gaps for fillings below $\nu=1$ than for fillings above $\nu=1$, in agreement with experiment \cite{CornellWigner,CaliforniaWigner,Berkeley,STM_Wigner, ContinuousWigner,FrustratedWignerMott}. We note however that remote band effects, neglected in our single-band study, are likely to play an equally important role in the particle-hole asymmetry seen in experiment. In appendix B we show the effects of including remote bands in our calculations, indicating that the importance of mixing with remote bands increases with filling factor. This is in agreement with different experimental measures where states for smaller hole fillings of the topmost band always have larger charge gaps \cite{CornellWigner,CaliforniaWigner,Berkeley,FrustratedWignerMott}.

Finally we comment on the importance of two items that can be
relevant in experiment, gate distance and disorder.  First, extended Coulomb interactions that trigger generalized Wigner crystal formation, are much more sensitive to the sample-gate coupling than the on-site interactions and can be controlled by varying the 
distance to the gate electrodes. The results presented in the main text have been obtained for the limit of infinite gate distance. In appendix B we study the effect of modifying the distance $d$ to gates on our phase diagrams for the fillings multiples of $\nu=1/3$, finding the same general trends with only quantitative changes. 

Unrotated bilayers with a lattice mismatch, as considered here,
eliminate the twist-variation source of disorder.  Disorder is always present 
however, and it may have some importance in determining the details of the 
observed metal-insulator transitions \cite{DisorderMITDebanjan, DisorderVlad, DisorderDasSarma} at fractional and half-filling. Our calculations which neglect
disorder nevertheless reproduce many qualitative and quantitative experimental features.

In this work we have focused on generalized Wigner crystal states on a triangular moiré superlattice, appearing on TMD heterobilayers and tuned by the displacement field. Some TMD homobilayers can be approximated by honeycomb moiré superlattice models \cite{Mattia} and incompressible states are also expected to appear if long-range interactions are strong enough \cite{LiangFuMonteCarlo,HoneycombWCTenessee}. In that case applying a displacement field breaks inversion symmetry and induces a complex hopping, which could modify the picture presented here and deserves further analysis. In a broader context, recent experimental efforts on designing patterned dielectrics that induce a superlattice on semiconductors or semimetals \cite{PatternedDean} could also host generalized Wigner crystals and the results obtained here would apply in the triangular case of those systems.
\newline

{\it Acknowledgments} -- We thank Nai Chao Hu, Eun-Ah Kim, Johannes Motruk and Yiqing Zhou for helpful interactions. This work was supported by the U.S. Department of Energy, Office of Science, Basic Energy Sciences, under Awards $\#$ DE-SC0022106. The authors acknowledge the Texas Advanced Computing Center (TACC) at The University of Texas at Austin for providing high-performance computer resources. PP acknowledges support from the Polish National Science Centre based on Decision No. 2021/41/B/ST3/03322.

\bibliography{refs}

\onecolumngrid
\appendix
\section{Structure factors of the classical crystal states}
%
The Brillouin zone mesh of size $M=9$ includes the ${\bm \gamma}$--point, the ${\bm \kappa}/{\bm \kappa}^{\prime}$--points and six internal points that we label ${\bm \vartheta}$. We want to calculate the values that the structure factor acquires at these points for the magnetic states presented in the main text, but in the classical limit and for a lattice with an arbitrary number of sites $M$. The coordinates of the momentum points of interest are given by ${\bm \gamma}=k_{\theta}(0,0)$, ${\bm \vartheta}=k_{\theta}(1/3,0)$ (due to the symmetry, it suffices to take one of the six internal points) and ${\bm \kappa}/{\bm \kappa^{\prime}}=k_{\theta}(1/2,\pm1/2\sqrt{3})$, with $k_{\theta}=4\pi/\sqrt{3}a_M$. For convenience, we will consider a triangular lattice with $M$ sites, where $M$ is a multiple of 3, such that we can divide the lattice into three triangular sublattices that we denote $A$, $B$ and $C$, each containing $M/3$ sites. We want to calculate 
\begin{align}
    \mathcal{S}({\bf q})=\frac{1}{M^2}\sum_{i,j}e^{ i\,{\bf q} \cdot({\bf R}_i-{\bf R}_j) } 
    \braket{{\bf S}_i\cdot {\bf S}_j},
\end{align}
For magnetic states in the classical limit, we can replace spin operators ${\bf S}_i$ by vectors of norm $1/2$, and expectation values $\langle {\bf S}_i\cdot {\bf S}_j\rangle$ become dot products.
\newline

For the classical state at $\nu=1/3$--filling, only one of the three sublattices is filled (we choose it to be the $A$--sublattice) and we can divide it again into three sublattices $A1$, $A2$ and $A3$ with different spin orientations, obtaining the Néel state. Their coordinates within the magnetic unit cell are ${\bf R}_{A1}=a_M^{\prime}(0,0)$, ${\bf R}_{A2}=a_M^{\prime}(1,\sqrt{3})$, ${\bf R}_{A3}=a_M^{\prime}(-1,\sqrt{3})$, with $a_M^{\prime}=\sqrt{3}a_M/2$. We note that ${\bf S}_i\cdot {\bf S}_j= 1/4$ when the two spins at $i$ and $j$ are aligned and that ${\bf S}_i\cdot {\bf S}_j=-1/8$ when spin orientations differ by $\pm 2\pi/3$. Therefore we get 
\begin{align}
     \mathcal{S}({\bm \gamma})&=\frac{1}{M^2}\left( \sum_{A1}\left[\sum_{A1}\left(\frac{1}{4}\right)+\sum_{A2}\left(-\frac{1}{8}\right)+\sum_{A3}\left(-\frac{1}{8}\right) \right]+\sum_{A2}\left[\sum_{A1}\left(-\frac{1}{8}\right)+\sum_{A2}\left(\frac{1}{4}\right)+\sum_{A3}\left(-\frac{1}{8}\right) \right]\right. \nonumber \\
     &\qquad \qquad \left. +\sum_{A3}\left[\sum_{A1}\left(-\frac{1}{8}\right)+\sum_{A2}\left(-\frac{1}{8}\right)+\sum_{A3}\left(\frac{1}{4}\right) \right]\right)=0.
\end{align}
For the edge points of the Brillouin zone the phase factors are $\text{exp} \left[i {\bm \kappa}\cdot({\bm R_{Ai}}-{\bm R_{Aj}})\right]=1$, with $i\neq j$, hence the calculation is similar to the ${\bm \gamma}$-case,
\begin{align}
    \mathcal{S}({\bm \kappa})=\mathcal{S}({\bm \kappa^{\prime}})=\frac{1}{M^2}\,3\,\sum_{A1}\left( \frac{M}{9}\right)\left[\frac{1}{4}-\frac{1}{8}-\frac{1}{8} \right]=0.
\end{align}
For the internal point we obtain a finite value,
\begin{align}
    \mathcal{S}({\bm \vartheta})=\frac{1}{M^2}\, 3\,\sum_{A1}\left( \frac{M}{9}\right)\left[ \frac{1}{4}+\left(-\frac{1}{8}\right)e^{i\frac{2\pi}{3}}+\left(-\frac{1}{8}\right)e^{-i\frac{2\pi}{3}} \right]=\frac{1}{M^2}\left( \frac{M}{3}\right)\left(\frac{M}{9} \right)\left(\frac{3}{8}\right)=\frac{1}{72}.
\end{align}
For the $\nu=2/3$--filling classical ground state, we choose to populate sublattices $B$ and $C$ within the magnetic unit cell, with coordinates ${\bf R}_B=a_M^{\prime}(0,2\sqrt{3})$ and ${\bf R}_C=a_M^{\prime}(0,4\sqrt{3})$, respectively. In this case the product is ${\bf S}_i\cdot {\bf S}_j=\pm1/4$, for aligned and anti-aligned spins respectively. The structure factor at the three points of interest is given by 
%
\begin{align}
    \mathcal{S}({\bm \gamma})=\frac{1}{M^2}\left[\sum_{B}\left(\sum_{B^{\prime}}\frac{1}{4}+\sum_{C}-\frac{1}{4} \right)+\sum_{C}\left(\sum_{B}-\frac{1}{4}+\sum_{C^{\prime}}\frac{1}{4} \right)\right]=0.
\end{align}
\begin{align}
    \mathcal{S}({\bm \kappa})=\mathcal{S}({\bm \kappa^{\prime}})=\frac{1}{M^2}\left[ \sum_{B}\left( \sum_{B^{\prime}}\frac{1}{4}+\sum_{C}-\frac{1}{4}e^{-i\frac{2\pi}{3}}\right)+\sum_{C}\left( \sum_{B}-\frac{1}{4}e^{i\frac{2\pi}{3}}+\sum_{C^{\prime}}\frac{1}{4}\right) \right]=\frac{1}{M^2}\left( \frac{M}{3}\right)^2\left(\frac{3}{4} \right)=\frac{1}{12}.
\end{align}
\begin{align}
    \mathcal{S}({\bm \vartheta})&=\frac{1}{M^2}\left[ \sum_{B}\left( \sum_{B^{\prime}}\frac{1}{4}e^{i{\bm \vartheta}\cdot({\bf R}_B-{\bf R}_{B^{\prime}})}+\sum_{C}-\frac{1}{4}e^{i{\bm \vartheta}\cdot({\bf R}_B-{\bf R}_{C})}\right)+\sum_{C}\left(\sum_{B}\frac{1}{4}e^{i{\bm \vartheta}\cdot({\bf R}_C-{\bf R}_{B})}+\sum_{C^{\prime}}-\frac{1}{4}e^{i{\bm \vartheta}\cdot({\bf R}_C-{\bf R}_{C^{\prime}})} \right)\right]\nonumber \\
    &=\frac{1}{M^2}\left[ \sum_{B}\left(\frac{1}{4}-\frac{1}{4}\right)\left(\sum_{B^{\prime}}e^{i{\bm \vartheta}\cdot {\bf R}_{B^{\prime}}} \right)+\sum_{C}\left(\frac{1}{4}-\frac{1}{4}\right)\left(\sum_{C^{\prime}}e^{i{\bm \vartheta}\cdot {\bf R}_{C^{\prime}}} \right)\right]=0,
\end{align}
where in the last line we used that ${\bm R}_C={\bm R}_B+{\bm R}_0$, with ${\bm R}_0=a_M^{\prime}(0,2\sqrt{3})$ the vector that separates the two sublattice sites within each unit cell, to rewrite the summations and that $\text{exp}[i{\bm \vartheta}\cdot {\bm R}_0]=1$. 

\section{Effect of remote bands, distance to gates and some results for other system sizes}
\begin{figure}[h]
\centering
\includegraphics[width=0.85\linewidth]{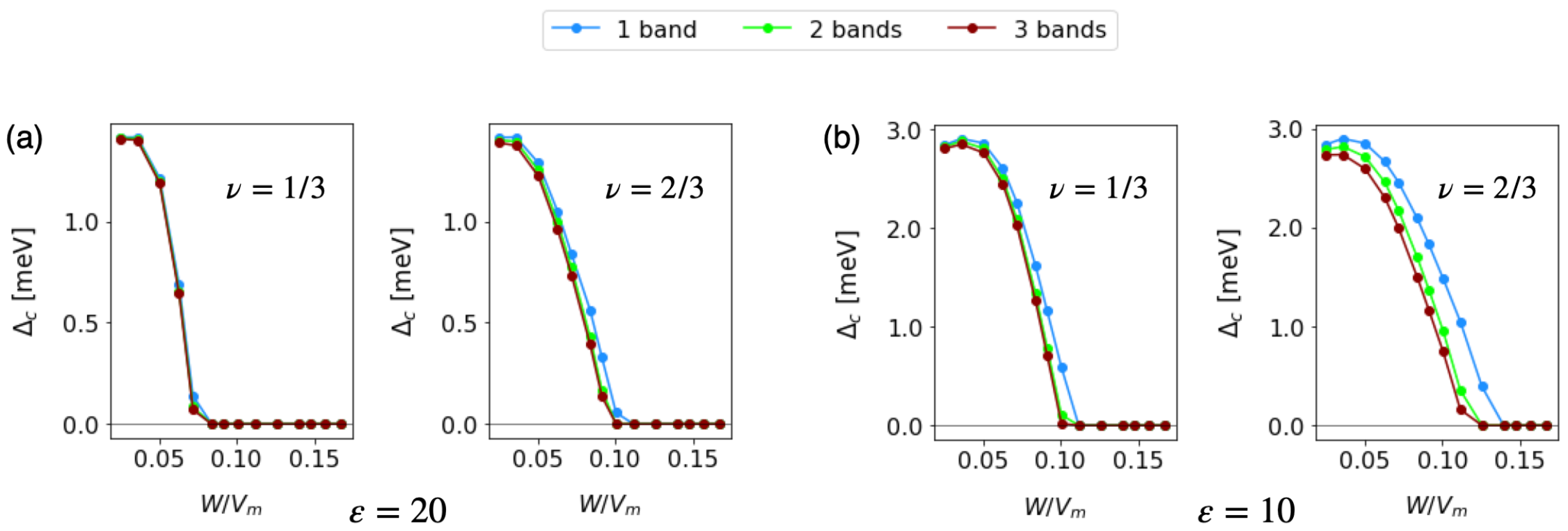}
\caption{Effect of remote bands in the evolution of charge gaps as a function of $W/V_m$ for fillings $\nu=1/3$ and $\nu=2/3$.}
\label{fig:Remote_Bands}
\end{figure}

The results presented in the main text correspond to projecting Coulomb interactions to the topmost moiré band, which is equivalent to mapping the system to an extended triangular Hubbard model. As we pointed out, this approximation is not always valid and is more expected to break down above half-filling $\nu>1$. In order to determine how much the inclusion of remote bands affects our results, in Fig. \ref{fig:Remote_Bands} we show charge gaps for $\nu=1/3$ and $\nu=2/3$, calculated after projecting the Hamiltonian to different numbers of bands in the ED calculation. 

Adding the remote bands has the effect of enlarging the Hilbert space, lowering the ground state energy. The phase boundary between antiferromagnetic and ferromagnetic regions at $\nu=2/3$--filling will be shifted because the remote bands effectively screen the dielectric function. Besides these quantitative changes, the general trends regarding the evolution of the charge gaps and the magnetic properties remain the same. From Fig. \ref{fig:Remote_Bands} we see that the effects of remote bands become more prominent for smaller dielectric constant $\varepsilon$ and higher filling factor $\nu$. This makes sense, since a small $\varepsilon$ means stronger Coulomb interactions, which will allow for virtual transitions to remote bands.
\newline
\begin{figure}[h]
\centering
\includegraphics[width=0.6\linewidth]{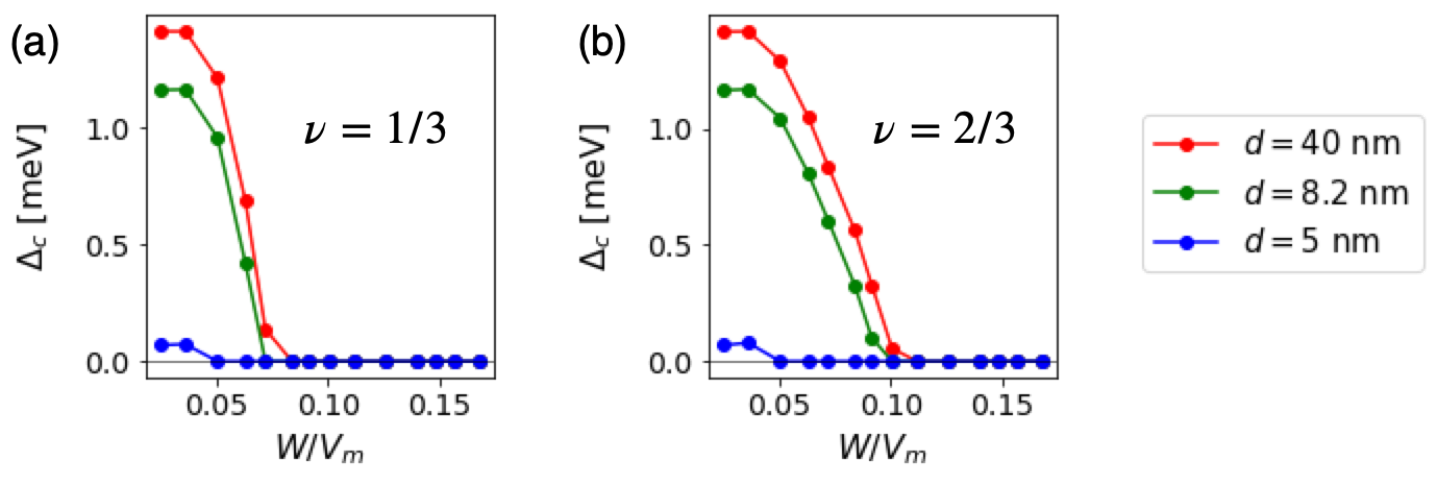}
\caption{Charge gaps as function of $W/V_m$ for (a) $\nu=1/3$--filling and (b) $\nu=2/3$--filling corresponding to different distances to gates and for $\varepsilon=20$.}
\label{fig:Gates}
\end{figure}

In Fig. \ref{fig:Gates} we show the effect of modifying gate distance on the charge gaps at fillings $\nu=1/3$ and $\nu=2/3$. The effect of the gates is introduced by modifying the interaction elements, Eq. \eqref{InteractionME}, to 
\begin{align}
\label{InteractionGates}
    V_{i,j,k,l}^{\sigma,\sigma^{\prime}}=\frac{1}{A}\sum_{\substack{{\bm G}_i,{\bm G}_j\\{\bm G}_k,{\bm G}_l}}\left(z^{*}_{{\bm k}_i,{\bm G}_i} z^{*}_{{\bm k}_j,{\bm G}_j}z_{{\bm k}_k,{\bm G}_k}z_{{\bm k}_l,{\bm G}_l}\right) \frac{2\pi e^2}{\varepsilon\,  q}
    \tanh \left( q\,d\right),
\end{align}
where $d$ is the distance from gates to sample, as shown in Fig. \ref{fig:SingleParticle}(a). Making the gates closer will screen the long-range interactions, which in turn will decrease the charge gap values. In particular, making the gate sufficiently close to the sample will cause the Wigner crystal states to completely disappear, as has been seen in experiments done in high proximity to gates \cite{FrustratedWignerMott}.
\newline
\begin{figure}[h]
\centering
\includegraphics[width=0.8\linewidth]{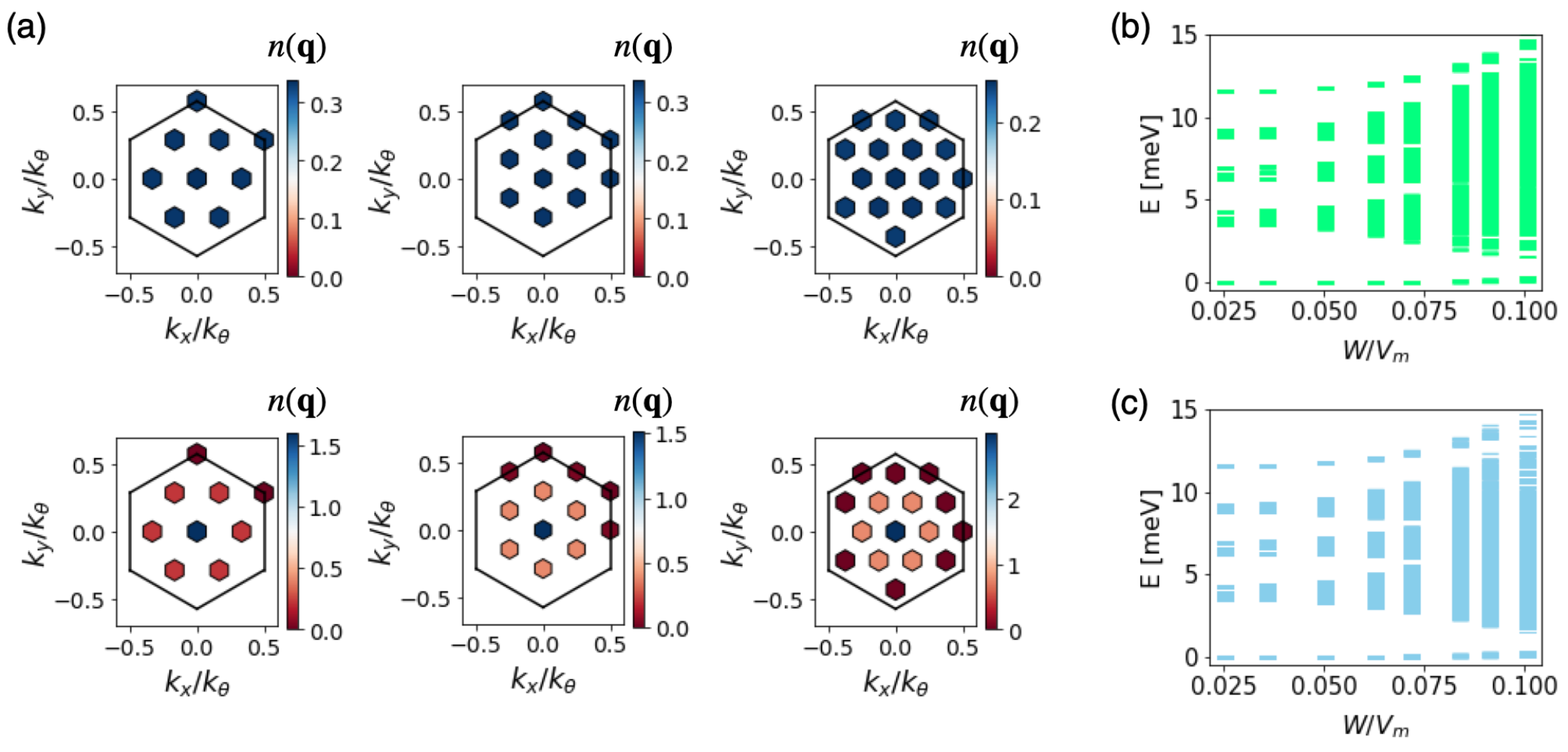}
\caption{(a) Ground state occupations in the atomic limit (top panel) and in the metallic regime (bottom panel). Results for system sizes $M=9$ and $M=12$ correspond to $\nu=1/3$ and results for $M=16$ correspond to filling $\nu=1/4$. (b) Many-body low-energy spectrum for $\nu=1/3$ obtained at size $M=12$, as a function of $W/V_m$. (c) Many-body low-energy spectrum for $\nu=2/3$ obtained at size $M=12$, as a function of $W/V_m$. These calculations were done for $\varepsilon=20$.}
\label{fig:Othersizes}
\end{figure}

Finally, in Fig. \ref{fig:Othersizes}(a), we show examples of occupation distributions for different system sizes in the atomic limit (top) and in the metallic limit (bottom). The top panel shows constant occupations for all system sizes, which are in agreement with an insulating state of localized spins. In the bottom panel, all system sizes present vanishing occupations at the points closer to the edge of the Brillouin zone. This is in agreement with a metallic state and the formation of a Fermi surface. 
\newline

The low-energy spectra for the $M=12$ system size are also shown in Fig. \ref{fig:Othersizes}(b) for $\nu=1/3$--filling and in Fig. \ref{fig:Othersizes}(c) for $\nu=2/3$--filling, as a function of $W/V_m$. We see the evolution of the ground state manifold, as well as the second (particle-hole excitation) and third (charged excitation) bands that we see in $M=9$. We additionally see two higher energy bands above the unbound particle-hole band for this system size. 

\end{document}